\documentclass[11pt,twoside]{article}

\usepackage{asp2006}
\usepackage{epsf}
\usepackage{psfig} \usepackage{lscape}
\usepackage{graphicx}

\begin{document}

\title{Clustering, host galaxies, and evolution of AGN}   
\author{Ryan C.\ Hickox and the Bo\"{o}tes\ survey collaboration}   
\affil{Harvard-Smithsonian Center for Astrophysics, 60 Garden Street, Cambridge, MA  02138} 

\begin{abstract} 
We explore the connection between different classes of active galactic
nuclei (AGNs) and the evolution of their host galaxies, by deriving
host galaxy properties, clustering, and Eddington ratios of AGNs
selected in the radio, X-ray, and infrared (IR) wavebands from the
wide-field (9 deg$^2$) Bo\"{o}tes\ survey.  We study a sample of 585
AGNs at $0.25<z<0.8$ using redshifts from the AGN and Galaxy Evolution
Survey (AGES). We find that radio and X-ray AGNs reside in relatively
large dark matter halos ($M_{\rm halo}\sim 3\times10^{13}$ and
$10^{13}$ $h^{-1}$ $M_{\sun}$, respectively) and are found in galaxies
with red and ``green'' colors. In contrast, IR AGNs are in less
luminous galaxies, have higher Eddington ratios, and reside in halos
with $M_{\rm halo}<10^{12}$ $M_{\sun}$.  We interpret these results in
terms of a general picture for AGNs and galaxy evolution, in which
quasar activity is triggered when $M_{\rm halo}\sim
10^{12}$--$10^{13}$ $M_{\sun}$, after which star formation ceases and AGN
accretion shifts to optically-faint, X-ray and radio-bright modes.

\end{abstract}


\section{Introduction}   
It is increasingly clear that a role in galaxy evolution is played by
active galactic nuclei (AGNs), as suggested by the tight correlation
between supermassive black hole (SMBH) mass and galaxy bulge
properties \citep[e.g.,][]{hickox_mago98, hickox_ferr00,
hickox_gebh00}, and the ability of AGN feedback to regulate star
formation in galaxy formation models \citep[e.g.,][]{hickox_bowe06gal,
hickox_hopk06apjs, hickox_crot06,hickox_khal08feedback,
hickox_bowe08flip}. Different modes of accretion may play varying
roles in galaxy evolution \citep[e.g.,][]{hickox_merl08agnsynth,
hickox_kauf08modes}, and observational clues to the links between AGN
and galaxies come from the host properties and dark matter halo masses
of different classes of AGNs.  In general, low-level accretion is
relatively bright in the radio and X-rays, while AGNs with higher
accretion rates are readily detected in the optical and IR.
Radio-selected AGNs are generally found in luminous red-sequence
galaxies in massive dark matter halos
\citep[e.g.,][]{hickox_mand08agnclust}, while X-ray AGNs reside in
galaxies with ``green'' colors and somewhat less massive halos
\citep[e.g.,][]{hickox_silv08host, hickox_gill08xcosmos}.  Optical
quasars are found in halos of mass $\sim 3\times 10^{12}$ $M_{\sun}$
at all redshifts \citep[e.g.,][]{hickox_croo05, hickox_shen07clust}.
These results indicate that the different modes of accretion occur at
different phases in galaxy evolution.  Here we present a study of the
host galaxies, clustering, and Eddington ratios for three classes of
AGNs (radio, X-ray, and IR-selected) in the 9 deg$^2$ Bo\"{o}tes\
field, and compare our results to a simple picture of AGN and galaxy
evolution.  We use a cosmology with $\Omega_{\rm m}=0.3$,
$\Omega_{\Lambda}=0.7$, $h=0.7$, and $\sigma_8=0.9$; the observations
and results are described in detail in \citet{hickox_hick09corrweb}.

\begin{figure}[t!]
\plottwo{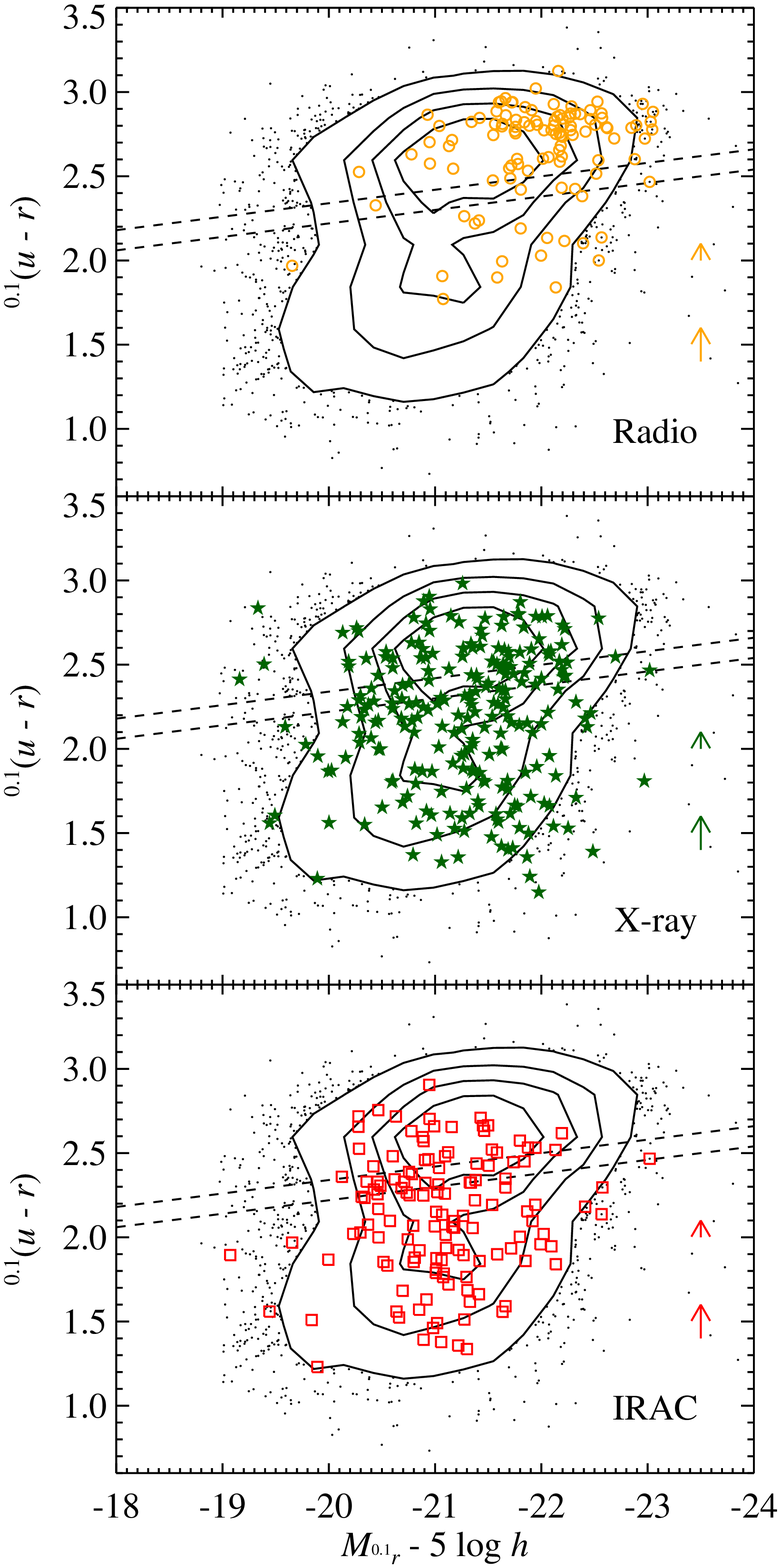}{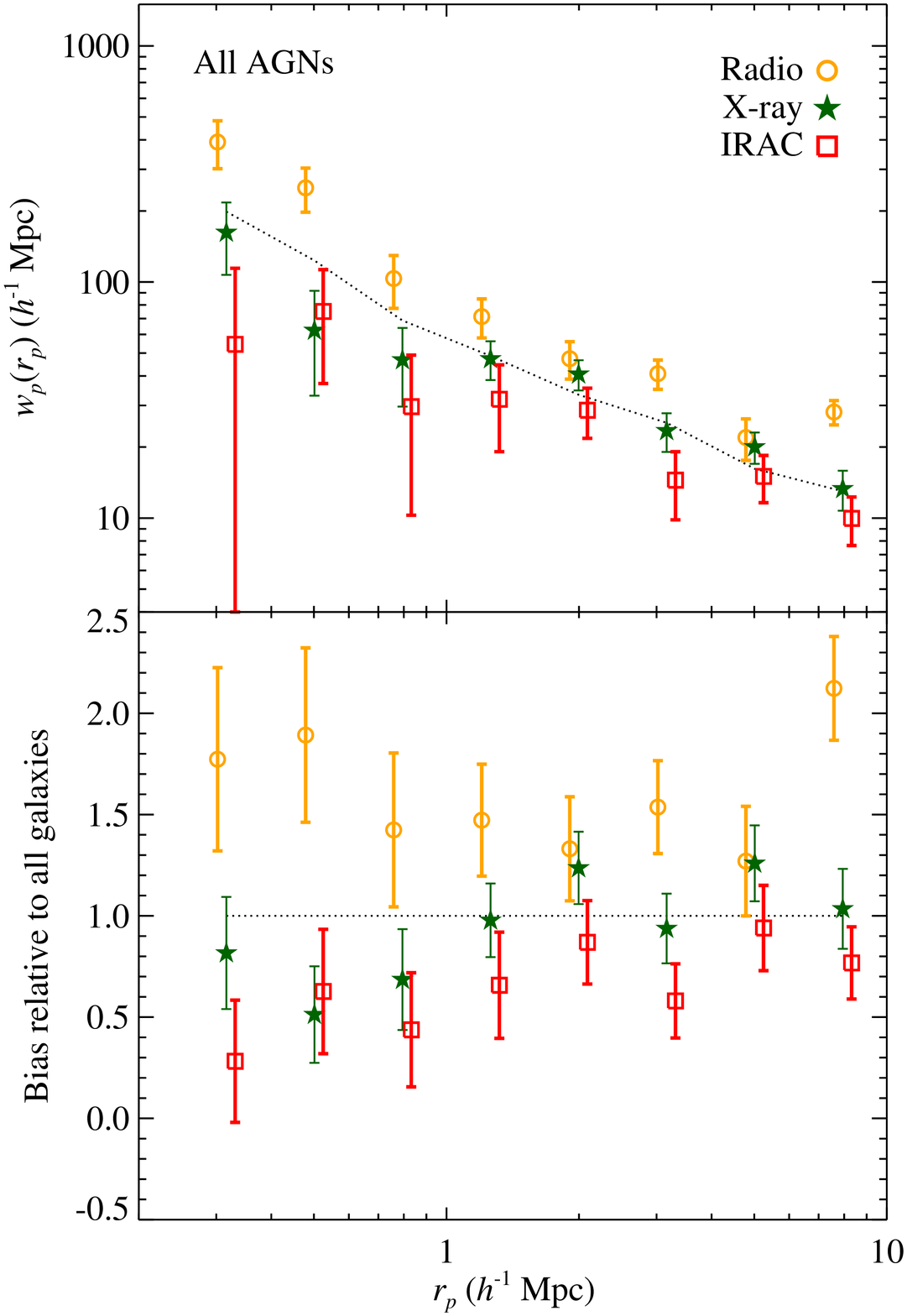}
\caption{{\em Left:} Optical colors and absolute magnitudes of AGNs
with extended optical counterparts.  Contours and black points show
the distribution for AGES galaxies, and dashed lines separate the blue
cloud from the red sequence.  Orange circles, green stars, and red
squares show radio, X-ray, and IR AGNs, respectively.  Arrows show
typical (small) corrections for nuclear contamination. {\em Right:}
The top panel shows the projected two-point cross-correlations of AGNs
with respect to all AGES galaxies (symbols as at left).  For
comparison, the dotted lines show the autocorrelation of AGES
galaxies. The bottom panel shows  bias for AGNs relative to AGES galaxies.}
\end{figure}

\section{Observations and AGN sample}
We use data from the 9 deg$^2$ Bo\"{o}tes\ multiwavelength survey,
with optical photometry from the NOAO Deep Wide-Field Survey
\citep{hickox_jann99}, X-ray data from the {\em Chandra} XBo\"{o}tes\
survey \citep{hickox_murr05}, {\em Spitzer} IRAC data from the IRAC
Shallow Survey \citep{hickox_eise04}, radio data from the Westerbork
Radio Telescope 1.4 GHz radio survey \citep{hickox_devr02}, and
redshifts from MMT/AGES (C.\ Kochanek et~al. 2009, in preparation).
We select 6262 galaxies and 585 AGNs in the redshift interval
$0.25<z<0.8$.  We select 122 radio AGNs with $P_{1.4\;
\rm{GHz}}>6\times10^{23}$ W Hz$^{-1}$, 362 X-ray AGNs with
$L_X>10^{42}$ ergs s$^{-1}$, and 238 IR AGNs using the color-color
criterion of \citet{hickox_ster05}.  Only 19 of the radio AGNs are
selected in either X-rays or the IR, although there is $\approx
30$--50\% overlap between the X-ray and IR samples \citep[for details
on this overlap sample see][]{hickox_hick09corrweb}.

\section{AGN host galaxies, clustering, and Eddington ratios}
We first derive host galaxy colors and luminosities (Fig.~1, {\em
left}) for AGNs with extended optical counterparts (for which the
nucleus is optically faint or obscured).  We include a small color
correction ($<0.2$ mag) to account for residual contamination from the
nucleus.  We find that radio AGNs mainly reside in luminous red
galaxies, while X-ray AGNs have a significant peak in the ``green
valley'' of the color distribution, as seen in other surveys
\citep[e.g.,][]{hickox_sanc04agnhost, hickox_nand07host}.  The IR AGNs
have similar colors to the X-ray AGNs, but with a smaller ``green''
peak and with lower average luminosities.

\begin{figure}[t!]
\begin{center}
\includegraphics[width=3in]{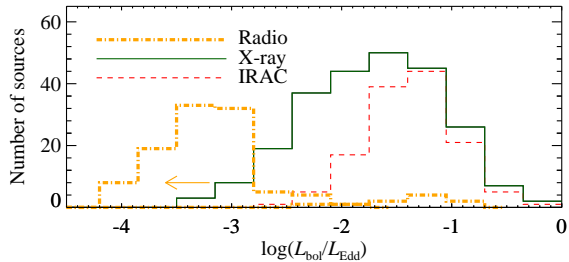}
\end{center}
\caption{Distributions in Eddington ratios for AGNs with extended
optical counterparts.  The $L_{\rm bol}$ and $L_{\rm bol}/L_{\rm Edd}$
estimates for the X-ray undetected radio AGNs show only upper limits
derived from X-ray stacking.  Radio, X-ray, and IR AGNs have
progressively higher typical Eddington ratios.}
\end{figure}

We next derive characteristic halo masses of the different AGN types
(for all AGNs, including optically-unresolved sources) by measuring
the projected two-point cross-correlation function [$w_p(r_p)$]
between AGNs and AGES galaxies on scales of $0.3<r_p<10$ $h^{-1}$ Mpc
(Fig.~1, {\em right}). Fitting a 3-D cross-correlation function of the
form $\xi=(r/r_0)^{-\gamma}$, we obtain $(r_0$ [$h^{-1}$ Mpc],
$\gamma)=(6.3\pm0.6, 1.8\pm0.2)$ for radio AGNs, $(4.7\pm0.3,
1.6\pm0.1)$ for X-ray AGNs, and $(3.7\pm0.4, 1.5\pm0.1)$ for IR AGNs.
Using the autocorrelation of the AGES galaxies and a model of dark
matter clustering \citep{hickox_smit03dm}, we then derive the absolute bias
of the dark matter halos that host the AGNs, and convert the bias to
$M_{\rm halo}$ \citep{hickox_shet01halo}.  We obtain $\log{M_{\rm halo}
(h^{-1} M_{\sun})}=13.5^{+0.1}_{-0.2}$ for radio AGNs,
$12.9^{+0.2}_{-0.3}$ for X-ray AGNs, and $11.7^{+0.6}_{-1.5}$ for IR
AGNs.  

Finally, we derive Eddington ratios ($\lambda=L_{\rm bol}/L_{\rm
Edd}$) for the AGNs with extended optical counterparts (Fig.~2).  We
estimate $M_{\rm BH}$ from the $L_{\rm bulge}$--$M_{\rm BH}$ relation
\citep{hickox_marc03}, and derive $L_{\rm bol}$ using X-ray and IR bolometric
corrections \citep{hickox_hopk07qlf}.  Radio AGNs have very low Eddington
ratios ($\lambda \la 10^{-3}$), while X-ray AGNs have intermediate
ratios ($10^{-3}\la\lambda\la1$) and IR AGNs have high ratios
($\lambda \ga 0.01$).

\begin{figure}[t]
\begin{center}
\includegraphics[width=4.4 in,angle=-90]{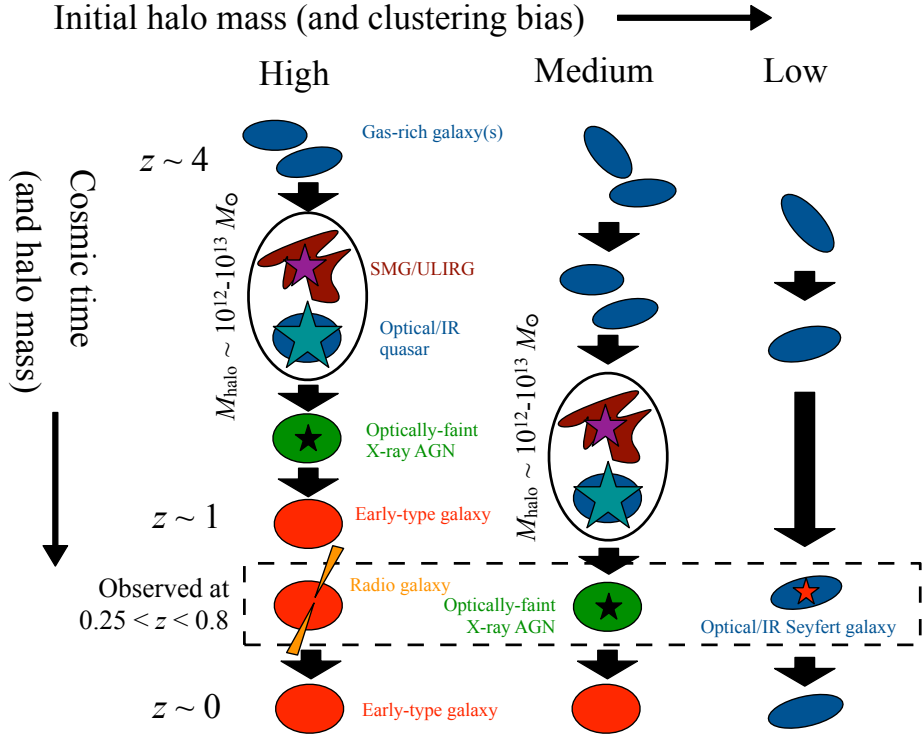}
\end{center}
\caption{\footnotesize Simple schematic of AGN and galaxy evolution,
as described in \S~4. Ovals represent the quasar/starburst
phase when $M_{\rm halo}\sim10^{12}$--$10^{13}$ $M_{\sun}$. The dashed
box shows the classes of AGNs that are observed at $0.25<z<0.8$.}
\end{figure}

\section{Evolutionary picture}
Our results are generally consistent with a simple picture of
SMBH and galaxy evolution (Fig.~3), motivated by recent observational
and theoretical studies.  As dark matter halos increase in mass
through the hierarchical growth of structure, galaxies initially form
as cold gas-rich, star-forming systems with rotationally-dominated dynamics.  When
the dark matter halo reaches a  mass $M_{\rm crit}\sim
10^{12}$--$10^{13}$ $M_{\sun}$ (shown by the ovals in Fig.~3), starburst
and luminous AGN (quasar) activity is fueled, for example by a major
galaxy merger or disk instabilities.  Subsequently, star
formation is quenched either by quasar feedback \citep{hickox_hopk06apjs}, or
by the creation of a virialized hot halo of gas that can no longer
efficiently cool \citep{hickox_deke06shock}.  At later times and higher
masses, lower-Eddington (X-ray and radio-bright) accretion
\citep[e.g.,][]{hickox_chur05smbh} may help prevent star formation
\citep[e.g.,][]{hickox_bowe06gal, hickox_crot06}.

  In this picture, our observed population of radio AGNs represents a
low-Eddington mode of accretion found in passive galaxies with
high-mass halos ($M_{\rm halo} > M_{\rm crit}$).  These galaxies have
already experienced the quasar phase and the quenching of star
formation.  The observed X-ray AGNs are in a somewhat higher-Eddington
phase that occurs in halos closer to $M_{\rm crit}$, during the
transition of the galaxy from the blue cloud to the red sequence,
while the observed IR-selected AGNs are the highest-Eddington
population, found in star-forming galaxies with small black holes and
dark matter halos that have not yet grown to $M_{\rm crit}$.  This
simple evolutionary picture naturally accounts for the observed  trends
in host galaxies, clustering, and Eddington ratios of the different
classes of AGNs.  In the future we will explore this scenario in more
detail, by comparing our results with detailed predictions of galaxy
formation models.

\acknowledgements 
R.C.H. was supported by {\em Chandra} grants
GO5-6130A and GO5-6121A.



\begin{thebibliography}{}
 

\bibitem[{{Bower} {et~al.}(2006){Bower}, {Benson}, {Malbon}, {Helly}, {Frenk},
  {Baugh}, {Cole}, \& {Lacey}}]{hickox_bowe06gal}
{Bower}, R.~G., {Benson}, A.~J., {Malbon}, R., {Helly}, J.~C., {Frenk}, C.~S.,
  {Baugh}, C.~M., {Cole}, S., \& {Lacey}, C.~G. 2006, \mnras, 370, 645

\bibitem[{{Bower} {et~al.}(2008){Bower}, {McCarthy}, \& {Benson}}]{hickox_bowe08flip}
{Bower}, R.~G., {McCarthy}, I.~G., \& {Benson}, A.~J. 2008, \mnras, 390, 1399

\bibitem[{{Churazov} {et~al.}(2005){Churazov}, {Sazonov}, {Sunyaev}, {Forman},
  {Jones}, \& {B{\"o}hringer}}]{hickox_chur05smbh}
{Churazov}, E., {Sazonov}, S., {Sunyaev}, R., {Forman}, W., {Jones}, C., \&
  {B{\"o}hringer}, H. 2005, \mnras, 363, L91

\bibitem[{{Croom} {et~al.}(2005){Croom}, {Boyle}, {Shanks}, {Smith}, {Miller},
  {Outram}, {Loaring}, {Hoyle}, \& {da {\^A}ngela}}]{hickox_croo05}
{Croom}, S.~M., et~al.\ 2005,  \mnras, 356, 415

\bibitem[{{Croton} {et~al.}(2006){Croton}, {Springel}, {White}, {De Lucia},
  {Frenk}, {Gao}, {Jenkins}, {Kauffmann}, {Navarro}, \& {Yoshida}}]{hickox_crot06}
{Croton}, D.~J., et~al.\ 2006, \mnras, 365, 11

\bibitem[{{de Vries} {et~al.}(2002){de Vries}, {Morganti}, {R{\" o}ttgering},
  {Vermeulen}, {van Breugel}, {Rengelink}, \& {Jarvis}}]{hickox_devr02}
{de Vries}, W.~H., {Morganti}, R., {R{\" o}ttgering}, H.~J.~A., {Vermeulen},
  R., {van Breugel}, W., {Rengelink}, R., \& {Jarvis}, M.~J. 2002, \aj, 123,
  1784

\bibitem[{{Dekel} \& {Birnboim}(2006)}]{hickox_deke06shock}
{Dekel}, A. \& {Birnboim}, Y. 2006, \mnras, 368, 2

\bibitem[{{Eisenhardt} {et~al.}(2004){Eisenhardt}, {Stern}, {Brodwin}, {Fazio},
  {Rieke}, {Rieke}, {Werner}, {Wright}, {Allen}, {Arendt}, {Ashby}, {Barmby},
  {Forrest}, {Hora}, {Huang}, {Huchra}, {Pahre}, {Pipher}, {Reach}, {Smith},
  {Stauffer}, {Wang}, {Willner}, {Brown}, {Dey}, {Jannuzi}, \&
  {Tiede}}]{hickox_eise04}
{Eisenhardt}, P.~R., et~al.\ 2004, \apjs, 154, 48

\bibitem[{{Ferrarese} \& {Merritt}(2000)}]{hickox_ferr00}
{Ferrarese}, L. \& {Merritt}, D. 2000, \apjl, 539, L9

\bibitem[{{Gebhardt} {et~al.}(2000){Gebhardt}, {Bender}, {Bower}, {Dressler},
  {Faber}, {Filippenko}, {Green}, {Grillmair}, {Ho}, {Kormendy}, {Lauer},
  {Magorrian}, {Pinkney}, {Richstone}, \& {Tremaine}}]{hickox_gebh00}
{Gebhardt}, K., et~al.\ 2000, \apjl, 539, L13

\bibitem[{{Gilli} {et~al.}(2008){Gilli}, {Zamorani}, {Miyaji}, {Silverman},
  {Brusa}, {Mainieri}, {Cappelluti}, {Daddi}, {Porciani}, {Pozzetti}, {Civano},
  {Comastri}, {Finoguenov}, {Fiore}, {Salvato}, {Vignali}, {Hasinger}, {Lilly},
  {Impey}, {Trump}, {Capak}, {McCracken}, {Scoville}, {Taniguchi}, {Carollo},
  {Contini}, {Kneib}, {Le Fevre}, {Renzini}, {Scodeggio}, {Bardelli},
  {Bolzonella}, {Bongiorno}, {Caputi}, {Cimatti}, {Coppa}, {Cucciati}, {de la
  Torre}, {de Ravel}, {Franzetti}, {Garilli}, {Iovino}, {Kampczyk}, {Knobel},
  {Kovac}, {Lamareille}, {Le Borgne}, {Le Brun}, {Maier}, {Mignoli}, {Pello'},
  {Peng}, {Perez Montero}, {Ricciardelli}, {Tanaka}, {Tasca}, {Tresse},
  {Vergani}, {Zucca}, {Abbas}, {Bottini}, {Cappi}, {Cassata}, {Fumana},
  {Guzzo}, {Leauthaud}, {Maccagni}, {Marinoni}, {Memeo}, {Meneux}, {Oesch},
  {Scaramella}, \& {Walcher}}]{hickox_gill08xcosmos}
{Gilli}, R., et~al.\ 2008, \aap\ in press (arXiv:0810.4769)

\bibitem[{{Hickox} {et~al.}(2009){Hickox}, {Jones}, {Forman}, {Murray},
  {Kochanek}, {Eisenstein}, {Jannuzi}, {Dey}, {Brown}, {Eisenhardt}, {Gorjian},
  {Brodwin}, {Narayan}, {Cool}, {Kenter}, {Caldwell}, \&
  Anderson}]{hickox_hick09corrweb}
{Hickox}, R.~C., et~al.\ 2009, \apj\ in press (arXiv:0901.4121)

\bibitem[{{Hopkins} {et~al.}(2006){Hopkins}, {Hernquist}, {Cox}, {Di Matteo},
  {Robertson}, \& {Springel}}]{hickox_hopk06apjs}
{Hopkins}, P.~F., {Hernquist}, L., {Cox}, T.~J., {Di Matteo}, T., {Robertson},
  B., \& {Springel}, V. 2006, \apjs, 163, 1

\bibitem[{{Hopkins} {et~al.}(2007){Hopkins}, {Richards}, \&
  {Hernquist}}]{hickox_hopk07qlf}
{Hopkins}, P.~F., {Richards}, G.~T., \& {Hernquist}, L. 2007, \apj, 654, 731

\bibitem[{{Jannuzi} \& {Dey}(1999)}]{hickox_jann99}
{Jannuzi}, B.~T. \& {Dey}, A. 1999, in ASP Conf.\ Ser.\ 191: Photometric
  Redshifts and the Detection of High Redshift Galaxies, ed. R.~{Weymann},
  L.~{Storrie-Lombardi}, M.~{Sawicki}, \& R.~{Brunner} (San Francisco: ASP),
  111

\bibitem[{{Kauffmann} \& {Heckman}(2008)}]{hickox_kauf08modes}
{Kauffmann}, G. \& {Heckman}, T.~M. 2008, submitted to \mnras\
  (arXiv:0812.1224)

\bibitem[{{Khalatyan} {et~al.}(2008){Khalatyan}, {Cattaneo}, {Schramm},
  {Gottl{\"o}ber}, {Steinmetz}, \& {Wisotzki}}]{hickox_khal08feedback}
{Khalatyan}, A., {Cattaneo}, A., {Schramm}, M., {Gottl{\"o}ber}, S.,
  {Steinmetz}, M., \& {Wisotzki}, L. 2008, \mnras, 387, 13

\bibitem[{{Magorrian} {et~al.}(1998){Magorrian}, {Tremaine}, {Richstone},
  {Bender}, {Bower}, {Dressler}, {Faber}, {Gebhardt}, {Green}, {Grillmair},
  {Kormendy}, \& {Lauer}}]{hickox_mago98}
{Magorrian}, J., et~al.\ 1998, \aj, 115, 2285

\bibitem[{{Mandelbaum} {et~al.}(2008){Mandelbaum}, {Li}, {Kauffmann}, \&
  {White}}]{hickox_mand08agnclust}
{Mandelbaum}, R., {Li}, C., {Kauffmann}, G., \& {White}, S.~D.~M. 2008,
  submitted to \mnras\ (arXiv:0806.4089)

\bibitem[{{Marconi} \& {Hunt}(2003)}]{hickox_marc03}
{Marconi}, A. \& {Hunt}, L.~K. 2003, \apjl, 589, L21

\bibitem[{{Merloni} \& {Heinz}(2008)}]{hickox_merl08agnsynth}
{Merloni}, A. \& {Heinz}, S. 2008, \mnras, 388, 1011

\bibitem[{{Murray} {et~al.}(2005){Murray}, {Kenter}, {Forman}, {Jones},
  {Green}, {Kochanek}, {Vikhlinin}, {Fabricant}, {Fazio}, {Brand}, {Brown},
  {Dey}, {Jannuzi}, {Najita}, {McNamara}, {Shields}, \& {Rieke}}]{hickox_murr05}
{Murray}, S.~S., et~al.\ 2005, \apjs, 161, 1

\bibitem[{{Nandra} {et~al.}(2007){Nandra}, {Georgakakis}, {Willmer}, {Cooper},
  {Croton}, {Davis}, {Faber}, {Koo}, {Laird}, \& {Newman}}]{hickox_nand07host}
{Nandra}, K., et~al.\ 2007, \apjl, 660, L11

\bibitem[{{S{\'a}nchez} {et~al.}(2004){S{\'a}nchez}, {Jahnke}, {Wisotzki},
  {McIntosh}, {Bell}, {Barden}, {Beckwith}, {Borch}, {Caldwell},
  {H{\"a}ussler}, {Jogee}, {Meisenheimer}, {Peng}, {Rix}, {Somerville}, \&
  {Wolf}}]{hickox_sanc04agnhost}
{S{\'a}nchez}, S.~F., et~al.\ 2004, \apj, 614, 586

\bibitem[{{Shen} {et~al.}(2007){Shen}, {Strauss}, {Oguri}, {Hennawi}, {Fan},
  {Richards}, {Hall}, {Gunn}, {Schneider}, {Szalay}, {Thakar}, {Vanden Berk},
  {Anderson}, {Bahcall}, {Connolly}, \& {Knapp}}]{hickox_shen07clust}
{Shen}, Y., et~al.\ 2007, \aj, 133, 2222

\bibitem[{{Sheth} {et~al.}(2001){Sheth}, {Mo}, \& {Tormen}}]{hickox_shet01halo}
{Sheth}, R.~K., {Mo}, H.~J., \& {Tormen}, G. 2001, \mnras, 323, 1

\bibitem[{{Silverman} {et~al.}(2008){Silverman}, {Mainieri}, {Lehmer},
  {Alexander}, {Bauer}, {Bergeron}, {Brandt}, {Gilli}, {Hasinger}, {Schneider},
  {Tozzi}, {Vignali}, {Koekemoer}, {Miyaji}, {Popesso}, {Rosati}, \&
  {Szokoly}}]{hickox_silv08host}
{Silverman}, J.~D., et~al.\ 2008, \apj, 675, 1025

\bibitem[{{Smith} {et~al.}(2003){Smith}, {Peacock}, {Jenkins}, {White},
  {Frenk}, {Pearce}, {Thomas}, {Efstathiou}, \& {Couchman}}]{hickox_smit03dm}
{Smith}, R.~E., et~al.\ 2003, \mnras, 341, 1311

\bibitem[{{Stern} {et~al.}(2005){Stern}, {Eisenhardt}, {Gorjian}, {Kochanek},
  {Caldwell}, {Eisenstein}, {Brodwin}, {Brown}, {Cool}, {Dey}, {Green},
  {Jannuzi}, {Murray}, {Pahre}, \& {Willner}}]{hickox_ster05}
{Stern}, D., et~al.\ 2005, \apj, 631, 163


\end{thebibliography}
\end{document}